# Multifaceted Accretion: The Interplay of Turbulence, Resistivity, Thermal Transport, and Dust around Black Holes

Asish Jyoti Boruah[1], Liza Devi[1], Biplob Sarkar[1]
[1]Department of Applied Sciences, Tezpur University, Napaam, Assam-784028, India
*Email ID: biplobs@tezu.ac.in (Corresponding Author)

**Abstract**: Accretion near black holes (BHs) is multidimensional, with turbulence, resistivity, thermal transport, and dust dynamics all playing essential roles. In cold accretion discs (ADs) or the region of an AD where magnetic fields (MFs) are negligible (or absent), hydrodynamic (HD) turbulence is probably dominating. However, Magneto-rotational instability (MRI) is the primary cause of turbulence in ADs. Significant velocity variations and rapid pressure changes are characteristics of turbulent flows, which allow better mixing and more angular momentum (AM) and energy transfer. Also, in accretion flow (AF), the interaction between turbulence and resistivity determines the efficiency of energy dissipation and heat transfer. Radiation, convection, and thermal conduction (TC) are the heat transport modes existing in AFs, where TC enables energy transfer in accreting materials via heat flux. Moreover, convection, also generated by turbulence, significantly impacts the stability of the AD and its vertical structure. The disc may be affected by radiation from the AD surrounding the BH when X-ray emission occurs. The emission from the disc is also affected by dust particles. Dust grains near BH are exposed to high temperatures and intense radiation, which might affect the flow characteristics, as seen in Active Galactic Nuclei (AGN). This chapter highlights the combined effect of turbulence, resistivity, transport mechanisms, and dust particles on BH AF. Future studies in this field must thoroughly investigate how dust, transport mechanisms, and turbulence interact in the BH accretion system.

## 1. Introduction

Accretion refers to the process through which matter flows toward the center of a gravitating object. When the central object is a BH, the most spectacular accretion observations occur. The BH's event horizon (with a spherical radius of about $\sim \frac{GM}{c^2}$) draws surrounding matter, which is then swallowed (Bu & Zhang, 2024; Pringle & King, 2007). When matter spirals towards BH, we observe the generation of AD, which is a hot, swirling mass of dust and gas (Jyoti Boruah et al., 2024). Broadly, ADs can be divided into the following categories: (i) protostellar discs, the regions of stellar and planetary formation; (ii) discs formed by mass transfer in binary star systems, which are the heart of compact X-ray sources and eruptive novae; and (iii) discs in AGN, which are believed to be the universe's most luminous sources (Balbus & Hawley, 1998; Das et al., 2021). In this study, our focus will mainly be on the last two categories. The multifaceted and intricate process of accretion onto BHs is essential to several high-energy astrophysical phenomena, such as the formation of stellar-mass BHs, AGNs, and ultra-luminous X-ray sources (ULXs). There are various processes that affect the AF and AD dynamics. Turbulence, resistivity, heat transport, and dynamics of dust interact extensively at the very heart of this accretion process.

For many years, BH accretion has been a vibrant field of study, and our comprehension of the underlying physical processes has advanced significantly. Many excellent reviews have been published over the years, which has considerably benefited our process of reviewing ADs. This gives us the advantage of being able to focus on the disc turbulence, dust, resistivity, thermal transport mechanisms, and their interplay thoroughly. Hence, in this chapter, the combined effects of those processes on BH AF will be thoroughly reviewed. Our discussion will focus on the present understanding of these physical processes and how they shape the observational characteristics of BH accretion systems.

## 2. Turbulence in AD

**Hydrodynamical turbulence:** One of the fundamental processes in the universe is the radial transport of AM in AD. It controls how ADs dynamically evolve and affects everything from planet formation to the development of supermassive BHs (Fromang & Lesur, 2019). Our understanding of hydrodynamical transport of AM in ADs has significantly increased in recent years due to a mix of experimental, theoretical, and numerical simulation (NS) studies (Sarkar et al., 2024). For several decades, there has been intense discussion about the existence and physical source of turbulence in ADs. MRI stops working or is less effective when the connection between MF and natural gas (in cold ADs) becomes weak, and that is when the investigation of pure HD origin of turbulence and transport phenomena comes into the picture (Mukhopadhyay et al., 2005). In general, an AD can become turbulent in two fundamental ways. In the first method, the flow exhibits some linear instability (LI), which eventually causes turbulence due to its nonlinear development. In the latter, the flow is linearly stable, and after a particular Reynolds number ($Re$) threshold is achieved, direct transitions from laminar to turbulent flows occur. Supercritical and (globally) subcritical are the terms used to describe the first and second types of transitions to turbulence (Lesur & Longaretti, 2005). Considering that ADs are shear flows and that shear flows are frequently destabilized and prone to turbulence in both the laboratory and daily life, in the late 90s and early 20s, a group of researchers concentrated on the idea that hydrodynamical turbulence might arise and result in improved and efficient transport. HD processes play a crucial role if we wish to arrive at a comprehensive and coherent description of AD dynamics, as new discoveries by various research groups have produced some surprising findings in recent years (Tevzadze et al., 2003; Richard & Zahn, 1999; Dubrulle et al., 2005; Fromang & Lesur, 2019).

**Supercritical Transition:** As critical values are reached, the base flow with LI transforms to a solution that is stable and steady, becoming progressively complex and eventually causing turbulence. For example, Taylor-Couette flow (TCF) and Rayleigh-Bénard convection flow with revolving inner cylinders. The fluid between concentric cylinders is driven by the azimuthal rotations of the inner and (or) outer walls of the cylinders in the TCF flow system. 2D roll cells known as a Taylor vortex (TV) appear steadily without circumferential variation or non-stationary fluctuation when TCF undergoes LI (Matsukawa & Tsukahara, 2025). Andereck et al. (1986) called this flow condition TV flow (TVF). The state of the flow changes to more complex states of wavy Taylor-vortex flow (WVF) and modulated wavy Taylor-vortex flow (MWV) as the inner-cylinder Reynolds number, $Re_{in}$, increases further. In 1923, Taylor first examined TCF's linear stability (Taylor, 1923).

**Subcritical Transition:** Plane Couette flows, and Poiseuille (pipe) flows are two examples of linearly stable flows. At high $Re,$ they become unstable and create turbulence. This type of transition is known as a subcritical transition and requires finite amplitude fluctuations rather than vanishingly small ones. In plane Couette studies, turbulence is readily recognized whenever $Re$ reaches approximately 1600, despite being linearly stable at all $Re$s' (Tillmark & Alfredsson, 1992; Fromang & Lesur, 2019). A long-standing mystery is the source of turbulence and hydrodynamical instability in Keplerian AD. The fluid motion within the Keplerian AD is linearly stable. Ghosh & Mukhopadhyay (2023) investigated how perturbation changes in this flow when a force is added. Such a force could be caused by feedback from outflows in ADs, the interactions of grain-fluid, thermal fluctuations (no matter how small), etc., and is thought to be characterized by stochastic behavior, like noise. By establishing the modified Landau equation and when Coriolis and external forces are present, the authors examined how nonlinear perturbations evolve. Even in the linear regime, the authors find that the otherwise least stable perturbation grows to a huge saturation amplitude with appropriate forcing and $Re$, resulting in nonlinearity and probable turbulence. In simple terms, forcing turns a linearly stable mode into an unstable one. In the presence of force, additionally, the authors demonstrate that perturbations nonlinearly diverge over a shorter period, resulting in an immediate transition to turbulence (Ghosh & Mukhopadhyay, 2023). Vertical shear instability (VSI), which is purely HD in nature, has received some attention in recent years because it only requires vertical shear in the angular velocity profile. The VSI arises as a natural outcome caused by irradiation from the central object due to a temperature gradient in the radial direction of the disc. NSs and linear analysis have demonstrated that the VSI performs efficiently for both vertically isothermal discs and fully radiative discs with stellar irradiation (Stoll & Kley, 2014; Nelson et al., 2013). In the well-known standard disc model by Shakura & Sunyaev (1973) (SS), the turbulence-related uncertainties are represented by a single constant ($\alpha$). This is an example of defining turbulence parameters and explaining the AM transport using an effective viscous prescription for many astrophysical applications (Stoll et al., 2017).

**MHD Turbulence:** The most probable cause of MHD turbulence in ADs is MRI. It is considered that the turbulence in saturated form, and, consequently, the rate of AM transport inside the disc are determined by viscosity (ν) and resistivity (η) (Fromang et al., 2009). Its origin has been the main focus of AD theories for many years. The MRI is now widely acknowledged as the most suitable choice (Balbus & Hawley, 1991). It saturates at amplitudes that match the observational results and causes MHD turbulence during its nonlinear phase. Moreover, it is considered that the turbulence in saturated form depends on both ν and η. Simulation studies indicate that ν increases with the magnetic Prandtl number ($P_m$), the ratio of ν to η. Without a mean field, Fromang et al. (2010) observed that at low $P_m$, MRI-induced MHD turbulence decreases (Fromang et al., 2009). They provided various strategies for monitoring the relative impact of numerical and explicit dissipation during local NSs of MRI-induced MHD turbulence. These methods allow for a systematic analysis of the saturated state of turbulence by varying dimensionless numbers. They observed a significant relationship between ν and magnetic $P_m$ but were unable to attain an asymptotic limit as ν and η decreased (Fromang et al., 2010). However, ordinary molecular ν orders of magnitude are too weak to account for the viscous astrophysical ADs. Rather, the MRI-induced turbulence is probably a result of the disc's stresses. Nevertheless, without the complexity of MHD, the stresses inside the disc may still be parametrized as an effective viscosity, and the typical HD machinery can be used. This is sometimes used since complete numerical treatments of MHD can be expensive, and analytical treatments can be pretty challenging to work with (Abramowicz & Fragile, 2013).

A basic mechanical model can be used to understand the MRI itself. Considering two gas particles joined by a line of MF and assembling them so that they are initially situated somewhat vertically apart but at the same cylindrical distance from the BH. Adding a small amount of extra AM to one of the particles (say the upper one) and subtracting a tiny bit from the lower one, results in the top particle moving outward to a new radius. Moving to a lesser radius, the bottom particle exhibits the reverse behavior. In the typical scenario, the upper particle will now orbit more slowly than the lower one. The MF line that connects these two particles will lengthen because of their different orbital speeds. The torque produced by the MF line's stretching creates extra stress. This stress shifts AM from the lower to the upper particle. It only strengthens the original perturbation, increasing the separation and AM transfer, which is the basic characteristic (Abramowicz & Fragile, 2013).

For many years, convection has been studied in the context of ADs as a way of transferring AM because it regulates the disc structures vertically by transferring heat. Convection may interact with the MRI in non-trivial ways, provided that a weak MF threads the gas and it is adequately ionized. When convection is assumed to be present, local MRI simulations with vertical stratification have recently shown significant variance in the AM transport, according to the ratio of stress to thermal pressure. The dynamic interaction between the two instabilities is unclear, even though MRI turbulence can act as a convectional heat source. The effects of ν and thermal diffusivity ($\chi$) in a non-ideal fluid minimises convection, and the Rayleigh number ($Ra$) measures the proportion of destabilizing to stabilizing processes: $Ra \equiv \frac{|N_B^2| H^4}{\nu \chi}$, where $H$ is the disc's scale height. Therefore, the requirement that the square of the buoyancy frequency be negative, or $N_B^2 < 0$, is a necessary but not sufficient condition in a viscid fluid. Both that $N_B^2 < 0$ and that the $Ra$ above a critical value $Ra_c$ are necessary for convective instability (CI), when both explicit ν and $\chi$ are considered (Held & Latter, 2021).

When viscous stress builds up in the disc as a result of viscosity changes or AM transfer, viscous instabilities develop (Devi et al., 2024). The eddying motions that characterize turbulence throughout a broad and continuous range of scales are caused and sustained by viscous instability (Mackay & Stephen, 2021).

## 3. Effect of Resistivity

The amount of flow distortion that results from MF dissipation depends on resistivity. The MF is a key component in the dynamic and complex process of magnetized plasma accretion onto BHs. The behavior of the AF is strongly dependent on the quantity of magnetic flux generated close to the event horizon (Nathanail et al., 2024). A dissipative mechanism essential to the structure and heating of hot AFs (HAFs) is resistivity (Ghoreyshi, 2020). The behavior of resistive HAFs with anisotropic pressure in situations where the MFs exhibit even symmetry in the z-axis around the midplane has been examined by Ghoreyshi & Khesali (2023). In the upper bound of the magnetic diffusivity parameter, their findings suggest that η may be a useful mechanism for heating HAFs. In BH ADs and jets, η plays a vital role in plasmoid growth, magnetic reconnection, and current sheet formation (Ripperda et al., 2019). Singh et al. (2021) used the PLUTO code's resistive MHD module to investigate 2D flow around the BH with low AM. They found that both the luminosity and the shock location change quasi-periodically along the equator for AFs with lower $\eta = 10^{-6}$ and 0.01. In contrast, the flow stabilizes and becomes symmetrical to the equator for higher values of $\eta = 0.1$ and 1.0 (Singh et al., 2021).

## 4. Thermal Transport Mechanisms

### 4.1. Conduction

Direct particle interactions transfer heat between one or more objects in contact. Conduction in ADs is caused by electron and ion collisions. The disc's conductivity depends on the plasma's temperature, density, and composition. The concept of conduction in ADs came into picture when Shapiro et al. (1976) presented a standard disc model and tried to explain bimodal behavior along with transition but could not be able to explain it. Later, Meyer & Meyer-Hofmeister (1994) presented their transition model for the slab corona model, where they revealed that the SS disc's (SSD) gas evaporates into the heated corona via TC. Honma (1996b) proposed a study on the global structure of bimodal ADs, where the author considered the diffusion of energy transport. According to Honma's assumption, when there is an entropy gradient in an ADAF, ν can be regarded as a source of turbulence in the flow and can be used to transport thermal energy (TE) (Honma, 1996a). Since ADAF has decreasing entropy in an outward direction, turbulent energy transport would carry TE to the outer disc. In the evaporation model of Meyer & Meyer-Hofmeister (1994), this radial transport of energy is caused by turbulence regulated analogous to TC. The extended work of Honma (1996a) has been done by Manmoto & Kato (2000) on the transition region in account of conduction. They found that in the radial direction, a large number of photons emit and travel through the ADAF, and they dominate the photons emitted from the transition region. So, through their work, we can conclude that there will always be significant radiative cooling, which is counterbalanced by turbulent heating (Manmoto & Kato, 2000). Depending on whether plasma is collisionless or collisional, we may get TC in two types. If the plasma in the flow is weakly collisional, then we can express the conduction in terms of heat flux, which is actually termed saturated TC (Tanaka & Menou, 2006; Bu et al., 2016; Mitra et al., 2023; Sarkar et al., 2025; Rezaie et al., 2025; Singh et al., 2025). Considering saturated TC, we can explain various properties (e.g., temperature, density, disc scale height, etc.) of HAF (Cowie & McKee, 1977). It is also very important to study TC from an observational point of view. Understanding the observed luminosity and spectral features of accreting systems requires a knowledge of how TC influences the accretion rate of compact objects. TC might affect the oscillations and variations that are observed (Pringle, 1981). The measured polarisation characteristics of X-ray, such as the location angle and degree of polarisation, can be affected by TC. Recently, numerous research groups have been working on mass loss via winds in sub-Eddington, radiatively inefficient AF considering MFs with parametric cooling. Through this, we can conclude that TC also affects the outflow properties.

### 4.2. Convection

Heat is transferred by convection, which occurs when fluids flow. When the disc is unstable to convective motions, convection occurs in ADs. This mechanism is often observable when the temperature gradient is sufficiently large to compensate for the stabilizing effects of rotation and MFs. The Convection-dominated Accretion Flow (CDAF) model is a theoretical framework for AF around compact objects that is based on convection. Instead of viscous heating, convective movements dominate the AF in CDAF. CIs may have a pivotal role in shaping the disc's structure, as noted by Begelman & Meier (1982). The authors anticipated that more heat would be transferred towards the disc surface because of the convection. Untrapped convective energy will propel a powerful wind that might ultimately collide with a jet. Later, Ryu & Goodman (1992) demonstrated that convection transports AM in thin ADs, provided the entropy in the AD is stratified vertically. Given a radially stratified medium, Narayan & Yi (1994) illustrated the validity of the argument of Begelman & Meier (1982). The authors argued that convection will almost exclusively occur in a radial direction because ADAF has entropy that increases inwards, making flows inherently unstable to CIs. Another possible cause of turbulent viscosity in ADAF is convection. By contrasting the convective and advective time-scales, Narayan & Yi (1995) discovered that convection only acts as a minor disturbance, changing the AF's characteristics somewhat without destroying the AF's overall structure. Convection can overwhelm advection and can propagate entropy outward more quickly than accretion can carry it in for an angle range of less than 30º around the rotating axis. Thus, we may state that convective motions will be a source of viscosity and will carry both energy and AM. Even if Balbus & Hawley (1991)'s instability were to fail, convective viscosity may continue the accretion.

When the ν is small ($\lesssim 0.1$), convection is high (Igumenshchev & Abramowicz, 1999; Igumenshchev et al., 2000). The structure of these CDAFs differs greatly from that of ADAFs, with mean radial velocity varying as $v \propto R^{-\frac{3}{2}}$ and accreting gas density varying as $\rho \propto R^{-\frac{1}{2}}$ (here, $R$ is the radius of the flow) (Igumenshchev & Abramowicz, 1999). In CDAFs, convection transports a luminosity $L_c \sim 10^{-3} - 10^{-2} \dot{M}c^2$ from small to large radii; the little mass accreting onto the BH provides the energy. According to Narayan et al. (2000) and Igumenshchev et al. (2000), this energy may be emitted as thermal bremsstrahlung from the CDAF's outer regions because flows with low ν ($< 0.1$) have a strong time-dependency. It has to do with the AF's increasing CI. As a result of this instability, heated convective bubbles form in the innermost region of the AF almost regularly. The period is determined by the viscous time scales. There is less viscosity when the duration is longer. Unstable matter motion on greater spatial scales is characteristic of flows with higher viscosities. We see erratic movements because low-viscosity flows are noisy and comprise many small-scale vortices. Viscosity's function in suppressing CI in AFs explains why there are two types of AFs with varying viscosities. The overall luminosity oscillates with a relative amplitude of $20 - 30\%$ due to the time variability of the low ν accretion fluxes. For flows with a ν value of 0.03–0.1, the resulting oscillations have frequencies in the range of around $1 - 20$ Hz. The measured quasi-periodic oscillations (QPOs) range of the galactic BH X-ray sources shows typical time-scales for this variability (Igumenshchev & Abramowicz, 1999).

**4.3. Radiation**

Heat transmission by electromagnetic (EM) waves is known as radiation. The interior portions of ADs surrounding BHs are predicted to be dominated by radiation and thermally unstable in the SS model when the rate of accretion exceeds a tiny fraction of Eddington, as shown by Shakura & Sunyaev (1976). It is crucial to understand the stability of radiation-dominated AD in order to grasp luminous BH sources. Considering the surface mass density ($\Sigma$) is constant, the origin of thermal instability (TI) may be most easily described in terms of the distinct dependency of the rate of cooling and heating on the temperature of the midplane (Piran, 1978). However, Since the idea that stress is related to radiation pressure (RP) is not proven to be true, the existence of TI has long been questioned. Later, Svensson & Zdziarski (1994) demonstrate that the disc itself may be maintained by gas pressure and TI would be eliminated if a significant portion of dissipation of local accretion power has occurred in optically thin regions (both above and below the disc). The turbulence observed in NSs of the nonlinear evolution of the MRI is now generally accepted to be similar to that observed in BH ADs (Balbus & Hawley, 1998). To explain the thermal stability of radiation-dominated discs, we contend that the dissipation of magnetic turbulence generates the stress. This stress provides the heat that is subsequently converted into RP. This allows us to explain how stress and pressure deduced from dimensional analysis are comparable, supporting the SS model and the estimation of instability. As a result, changes in magnetic energy cause changes in pressure, establishing a relationship between the two while preserving thermal stability (Hirose et al., 2009). The vertical structure of RP-supported ADs can be explored in spherical coordinates. In that case, we must consider the local energy balance per unit volume between cooling by advection, radiation, and viscous heating. Thus, there are primarily two sets of NSs on optically thick accretion processes. One is the use of a shearing box in 3D radiation magnetohydrodynamic (RMHD) simulations that concentrate on thin discs (Hirose et al., 2006; Krolik et al., 2007). The other is 2D simulations of global flows, such as thin discs and super-Eddington AFs, using radiation HD (RHD) or RMHD. At accretion rates near and beyond $\dot{M}_{Edd}$, the photosphere may be located at $\theta = 45°$, far from the equatorial plane. Additionally, Gu (2012) found that there will be a maximum accretion rate, which will help us comprehend ULX sources, the majority of which are not in a thermally dominant state. We can have three classes of BHs in ULXs: (i) normal stellar mass BHs ($\sim 10 M_\odot$), (ii) massive stellar BHs ($\sim 100 M_\odot$), and (iii) intermediate mass BHs ($10^2 - 10^4 M_\odot$). For most sources up to luminosities of $\sim$ a few $10^{40} \text{ergs}^{-1}$, a massive stellar BH with a modest super-Eddington rate of accretion appears to be responsible. The thin disc model predicts strong heat radiation from the disc, which is the traditional model for super-Eddington ADs. Nevertheless, observations have demonstrated that except for a small number of sources like M82 X-1 and HLX-1, ULXs are not in the thermally dominant condition (Feng & Kaaret, 2010; Davis et al., 2011). One way to conceptualize ULX radiation is as an optically thick disc with powerful outflows and $\dot{m} \lesssim \dot{m}_{max}$. Thermal radiation, which is often not dominating due to the modest $\dot{m}_{max}$, will be provided by the disc. Conversely, the outflows might contribute to the non-thermal radiation through the jet of the magnetically collimated outflow. They are driven by RP or by bulk motion Comptonization (Gu, 2012).

## 5. Effect of Dusts

In ADs, dust is essential to many astrophysical processes. The opacity and temperature structure of the disc is impacted by the scattered and absorbed photons of dust grains. The majority of the potential energy of the gas is released in AD's inner region. The accretion of gas powers the radiative output of an AGN (Bora et al., 2023). When dust is present, opacity increases 10–100 times more than when gas is the only matter present. This causes a significant increase in the coupling between gas and nucleus radiation (Dorodnitsyn & Kallman, 2021). The radius beyond which dust can exist creates a dust sublimation surface, which separates the outside portion, which is frequently connected to the dusty torus, from the inner, largely dust-free zone. For a $10^7 M_\odot$ BH, the virial theory indicates that the temperature of the obscuring gas should be $10^6$ K to be thick geometrically at a distance around, d~ 1.0 pc. This contradicts the obscurer's ability to sustain dust (Dorodnitsyn & Kallman, 2021; Krolik & Begelman, 1988). While Baskin & Laor (2018) calculated the opacity of the dust on the basis of the new data for grains of graphite, Czerny & Hryniewicz (2011) proposed that large-scale "failed" winds may be driven by the disc's pressure, local radiation on dust (Dorodnitsyn & Kallman, 2021). The disc is extremely slim because the dusty gas spiraling from galactic scales toward the nucleus is often rather cold. Gas gets warmer significantly closer to the BH because of internal viscous dissipation. Gas is warmer until the internally produced radiation begins to affect the disc's vertical structure by RP on dust grains. Since heated dust emits thermal energy, it impacts BH development and visible diagnostics. The accretion rate and burst time are influenced by dust mostly due to RP on dust, but dust opacity also decreases the extent of the ionized zone (Yajima et al., 2017; Toyouchi et al., 2019). By releasing radiation, dust contributes to the cooling of the disc and affects its thermal equilibrium. Further research is required because self-regulated accretion caused by the disc's own RP on dust may be crucial for controlling the expansion of supermassive BHs through accretion (Dorodnitsyn & Kallman, 2021).

## 6. Summary

BH AFs are complex systems in which various physical processes interact to influence dynamics and observable signatures. To understand these systems, one must understand how turbulence, resistivity, heat transport, and dust interact. Turbulence, for instance, is essential to AFs because it drives the transfer of AM and affects the stability of the flow. Resistivity impacts the dynamics of the flow's MF, which affects the generation of turbulence and overall stability. Thermal transport processes, including conduction, convection, and radiation, control the temperature and entropy of AFs. The AF's opacity, thermal transport, and chemical composition (such as metallicity) are all influenced by the interaction between the grains of dust and gas. The structure and evolution of the AF are determined by the interaction of these processes resulting in various observational signatures, including polarizations, spectral characteristics (Taverna et al., 2021), and variability (Nathanail et al., 2024).

Few studies have examined the thermodynamics that are directly related to observational data after decades of research on the dynamics of MRI turbulence. For example, it is still unclear what the thin disc transition radius actually is in relation to the observed features of X-ray binaries (e.g., the transition from hard to soft state), QPOs, and the breaking noise at high frequency, and what is the impact of viscous instabilities on variability, etc. (Q. Bu & Zhang, 2024)? The origin of high frequency QPOs, the origin and propagation of disc variability, degeneracy in spectral modeling, the location and structure of the coronae, and other unanswered questions remain. New instruments are needed to collect more photons and achieve higher sensitivity for fast variations. Also, numerous studies have investigated outflow characteristics in the context of TC, incorporating MF with parametric cooling. Instead of using parametric cooling, we may use physically driven cooling techniques like bremsstrahlung, synchrotron, and Comptonization to obtain more realistic insights. Observational parameters like local dissipative compactness of the coronae and spectral index can be estimated in conduction and radiation losses (Maciołek-Niedźwiecki et al., 1997). Since, convective motion cannot be well described by the axial symmetry assumptions of the latest model, future research on extreme HAFs, where convection has been challenging to illustrate in 3D structure, may reveal an equally challenging connection between MRI and convection (Held & Latter, 2021). Furthermore, the nature of the dusty AGN AD's vertical structure, the impact of incident central continuum on the structure, and the structure of the subsequent disc wind must all be resolved using RHD models (Baskin & Laor, 2018).

**Acknowledgements**

Biplob Sarkar acknowledges the financial support of UGC-BSR Start-Up-Grant for carrying out this work (No.F.30-558/2021 (BSR) dated-16/12/2021). The authors Asish Jyoti Boruah, Liza Devi, and Biplob Sarkar acknowledge the use of Meta AI and Google Gemini AI to improve the readability of the text. The authors also acknowledge the reviewer for thoroughly reviewing the manuscript and providing beneficial comments.

**Bibliography**

[1] Abramowicz, M. A., & Fragile, P. C. (2013). Foundations of Black Hole Accretion Disk Theory. *Living Reviews in Relativity*, *16*(1), 1. https://doi.org/10.12942/lrr-2013-1

[2] Andereck, C. D., Liu, S. S., & Swinney, H. L. (1986). Flow regimes in a circular Couette system with independently rotating cylinders. *Journal of Fluid Mechanics*, *164*, 155–183.

[3] Balbus, S. A., & Hawley, J. F. (1991). A powerful local shear instability in weakly magnetized disks. I-Linear analysis. II-Nonlinear evolution. *Astrophysical Journal, Part 1 (ISSN 0004-637X), Vol. 376, July 20, 1991, p. 214-233.*, *376*, 214–233.

[4] Balbus, S. A., & Hawley, J. F. (1998). Instability, turbulence, and enhanced transport in accretion disks. *Reviews of Modern Physics*, *70*(1), 1–53. https://doi.org/10.1103/RevModPhys.70.1

[5] Baskin, A., & Laor, A. (2018). Dust inflated accretion disc as the origin of the broad line region in active galactic nuclei. *Monthly Notices of the Royal Astronomical Society*, *474*(2), 1970–1994.

[6] Begelman, M. C., & Meier, D. L. (1982). Thick accretion disks-Self-similar, supercritical models. *Astrophysical Journal, Part 1, Vol. 253, Feb. 15, 1982, p. 873-896. Research Supported by the Science Research Council of England*, *253*, 873–896.

[7] Bora, H., Gogoi, N., Sarkar, B., Bhattacherjee, S., & Gogoi, R. (2023). Study of Properties of Active Galactic Nuclei and Observational Findings. In book: *Recent Trends in Physics Research (A Proceedings of XIII Biennial National Conference of Physics Academy of North East, PANE-2022) (ISBN: 978-93-90951-66-6), 209-214.*

[8] Bu, D.-F., Wu, M.-C., & Yuan, Y.-F. (2016). Effects of anisotropic thermal conduction on wind properties in hot accretion flow. *Monthly Notices of the Royal Astronomical Society*, *459*(1), 746–753.

[9] Bu, Q., & Zhang, S.-N. (2024). Black holes: Accretion processes in X-ray binaries. *Handbook of X-Ray and Gamma-Ray Astrophysics*, 3911–3938.

[10] Cowie, L. L., & McKee, C. F. (1977). The evaporation of spherical clouds in a hot gas. I-Classical and saturated mass loss rates. *Astrophysical Journal, Vol. 211, Jan. 1, 1977, Pt. 1, p. 135-146.*, *211*, 135–146.

[11] Czerny, B., & Hryniewicz, K. (2011). The origin of the broad line region in active galactic nuclei. *Astronomy & Astrophysics*, *525*, L8.

[12] Das, B., Sarkar, B., & Nath, A. (2021). Looking at the high-energy X-ray universe-An overview. *Journal of the Maharaja Sayajirao University of Baroda (ISSN: 0025-0422)*, Volume-55, No.2 2021, 224-231.

[13] Davis, S. W., Narayan, R., Zhu, Y., Barret, D., Farrell, S. A., Godet, O., Servillat, M., & Webb, N. A. (2011). The cool accretion disk in ESO 243-49 HLX-1: Further evidence of an intermediate-mass black hole. *The Astrophysical Journal*, *734*(2), 111.

[14] Devi, L., Boruah, A. J., & Sarkar, B. (2025). Accretion Disc Outbursts and Stability Analysis. *Physics Frontiers*, Vol.-I, pp: 8-11.

[15] Dorodnitsyn, A., & Kallman, T. (2021). A Physical Model for a Radiative, Convective Dusty Disk in AGN. *The Astrophysical Journal*, *910*(1), 67.

[16] Dubrulle, B., Marie, L., Normand, C., Richard, D., Hersant, F., & Zahn, J.-P. (2005). A hydrodynamic shear instability in stratified disks. *Astronomy & Astrophysics*, *429*(1), 1–13.

[17] Feng, H., & Kaaret, P. (2010). Identification of the X-ray Thermal Dominant State in an Ultraluminous X-ray Source in M82. *The Astrophysical Journal Letters*, *712*(2), L169.

[18] Fromang, S., & Lesur, G. (2019). Angular momentum transport in accretion disks: A hydrodynamical perspective. *EAS Publications Series*, *82*, 391–413.

[19] Fromang, S., Papaloizou, J., Lesur, G., & Heinemann, T. (2009). Numerical Simulations of MHD Turbulence in Accretion Disks. *Numerical Modeling of Space Plasma Flows: ASTRONUM-2008*, *406*, 9. https://adsabs.harvard.edu/full/2009ASPC..406....9F

[20] Fromang, S., Papaloizou, J., Lesur, G., & Heinemann, T. (2010). MHD turbulence in accretion disks: The importance of the magnetic Prandtl number. *EAS Publications Series*, *41*, 167–170.

[21] Ghoreyshi, S. M. (2020). Self-similar structure of resistive ADAFs with outflow and large-scale magnetic field. *Publications of the Astronomical Society of Australia*, *37*, e023.

[22] Ghoreyshi, S. M., & Khesali, A. (2023). The role of resistivity in hot accretion flows with anisotropic pressure: Comparing magnetic field models. *Publications of the Astronomical Society of Japan*, *75*(1), 52–70.

[23] Ghosh, S., & Mukhopadhyay, B. (2023). Hydrodynamical transport of angular momentum in accretion disks in the presence of nonlinear perturbations due to noise. *The Sixteenth Marcel Grossmann Meeting*, 295–306. https://doi.org/10.1142/9789811269776_0020

[24] Gu, W.-M. (2012). Radiation pressure-supported accretion disks: Vertical structure, energy advection, and convective stability. *The Astrophysical Journal*, *753*(2), 118.

[25] Held, L. E., & Latter, H. N. (2021). Magnetohydrodynamic convection in accretion discs. *Monthly Notices of the Royal Astronomical Society*, *504*(2), 2940–2960.

[26] Hirose, S., Krolik, J. H., & Blaes, O. (2009). Radiation-dominated disks are thermally stable. *The Astrophysical Journal*, *691*(1), 16.

[27] Hirose, S., Krolik, J. H., & Stone, J. M. (2006). Vertical structure of gas pressure-dominated accretion disks with local dissipation of turbulence and radiative transport. *The Astrophysical Journal*, *640*(2), 901.

[28] Honma, F. (1996a). Global structure and phase transition of advection-dominated accretion disks. *Basic Physics of Accretion Disks*, 31–36.

[29] Honma, F. (1996b). Global structure of bimodal accretion disks around a black hole. *Publications of the Astronomical Society of Japan*, *48*(1), 77–87.

[30] Igumenshchev, I. V., & Abramowicz, M. A. (1999). Rotating accretion flows around black holes: Convection and variability. *Monthly Notices of the Royal Astronomical Society*, *303*(2), 309–320.

[31] Igumenshchev, I. V., Abramowicz, M. A., & Narayan, R. (2000). Numerical simulations of convective accretion flows in three dimensions. *The Astrophysical Journal*, *537*(1), L27.

[32] Jyoti Boruah, A., Devi, L., & Sarkar, B. (2025). Relativistic Accretors and High Energy X-Ray view. *Physics Frontiers*, Vol.-I, pp: 33-40.

[33] Krolik, J. H., & Begelman, M. C. (1988). Molecular tori in Seyfert galaxies-Feeding the monster and hiding it. *Astrophysical Journal, Part 1 (ISSN 0004-637X), Vol. 329, June 15, 1988, p. 702-711. Research Supported by the Ball Corp., Rockwell International Corp., and Exxon Education Foundation.*, *329*, 702–711.

[34] Krolik, J. H., Hirose, S., & Blaes, O. (2007). Thermodynamics of an accretion disk annulus with comparable radiation and gas pressure. *The Astrophysical Journal*, *664*(2), 1045.

[35] Lesur, G., & Longaretti, P.-Y. (2005). On the relevance of subcritical hydrodynamic turbulence to accretion disk transport. *Astronomy & Astrophysics*, *444*(1), 25–44.

[36] Maciołek-Niedźwiecki, A., Krolik, J. H., & Zdziarski, A. A. (1997). Thermal conduction in accretion disk coronae. *The Astrophysical Journal*, *483*(1), 111.

[37] Mackay, E., & Stephen, K. (2021). The relevance of the linear stability theory to the simulation of unstable immiscible viscous-dominated displacements in porous media. *Journal of Petroleum Science and Engineering*, *207*, 109150.

[38] Manmoto, T., & Kato, S. (2000). Transition from standard disk to advection-dominated accretion flow. *The Astrophysical Journal*, *538*(1), 295.

[39] Matsukawa, Y., & Tsukahara, T. (2025). Switching between supercritical and subcritical turbulent transitions in inner cylinder rotating Taylor–Couette–Poiseuille flow. *International Journal of Heat and Fluid Flow*, *112*, 109667.

[40] Meyer, F., & Meyer-Hofmeister, E. (1994). Accretion disk evaporation by a coronal siphon flow. *Astronomy and Astrophysics (ISSN 0004-6361), Vol. 288, No. 1, p. 175-182*, *288*, 175–182.

[41] Mitra, S., Ghoreyshi, S. M., Mosallanezhad, A., Abbassi, S., & Das, S. (2023). Global transonic solution of hot accretion flow with thermal conduction. *Monthly Notices of the Royal Astronomical Society*, *523*(3), 4431–4440.

[42] Mukhopadhyay, B., Afshordi, N., and Narayan, R. (2005). Hydrodynamic Turbulence in Accretion Disks., *22nd Texas Symposium on Relativistic Astrophysics*, 465–470.

[43] Narayan, R., Igumenshchev, I. V., & Abramowicz, M. A. (2000). Self-similar accretion flows with convection. *The Astrophysical Journal*, *539*(2), 798.

[44] Narayan, R., & Yi, I. (1994). Advection-Dominated Accretion: A Self-Similar Solution. *The Astrophysical Journal*, *428*, L13. https://doi.org/10.1086/187381

[45] Narayan, R., & Yi, I. (1995). Advection-Dominated Accretion: Self-Similarity and Bipolar Outflows. *The Astrophysical Journal*, *444*, 231. https://doi.org/10.1086/175599

[46] Nathanail, A., Mizuno, Y., Contopoulos, I., Fromm, C. M., Cruz-Osorio, A., Moriyama, K., & Rezzolla, L. (2025). The impact of resistivity on the variability of black hole accretion flows. *Astronomy & Astrophysics*, *693*, A56.

[47] Nelson, R. P., Gressel, O., & Umurhan, O. M. (2013). Linear and non-linear evolution of the vertical shear instability in accretion discs. *Monthly Notices of the Royal Astronomical Society*, *435*(3), 2610–2632.

[48] Piran, T. (1978). The role of viscosity and cooling mechanisms in the stability of accretion disks. *Astrophysical Journal, Part 1, Vol. 221, Apr. 15, 1978, p. 652-660. Research Supported by the Science Research Council.*, *221*, 652–660.

[49] Pringle, J. E. (1981). Accretion discs in astrophysics. *In: Annual Review of Astronomy and Astrophysics. Volume 19.(A82-11551 02-90) Palo Alto, CA, Annual Reviews, Inc., 1981, p. 137-162.*, *19*, 137–162.

[50] Pringle, J. E., & King, A. (2007). *Astrophysical flows*. Cambridge University Press.

[51] Rezaie, S., Ghasemnezhad, M., & Golshani, M. (2025). Global Transonic Solution of magnetized dissipative accretion flow around non-rotating black holes with thermal

conduction. *New Astronomy*, vol. 116, Art. no. 102348, Elsevier. doi:10.1016/j.newast.2024.102348.

[52] Richard, D., & Zahn, J.-P. (1999). Turbulence in differentially rotating flows. What can be learned from the Couette-Taylor experiment. *Astronomy and Astrophysics*, *Vol. 347*, pp. 734–738.

[53] Ripperda, B., Bacchini, F., Porth, O., Most, E. R., Olivares, H., Nathanail, A., Rezzolla, L., Teunissen, J., & Keppens, R. (2019). General-relativistic resistive magnetohydrodynamics with robust primitive-variable recovery for accretion disk simulations. *The Astrophysical Journal Supplement Series*, *244*(1), 10.

[54] Ryu, D., & Goodman, J. (1992). Convective instability in differentially rotating disks. *The Astrophysical Journal, Vol. 388,* IOP, p. 438. doi:10.1086/171165.

[55] Sarkar, B., Dihingia, I. K., & Misra, R. (2025) Thermal conduction and thermal-driven winds in magnetized viscous accretion disk dynamics". *New Astronomy*, vol. 118, Art. no. 102377, Elsevier. doi:10.1016/j.newast.2025.102377.

[56] Sarkar, B., Devi, L., & Boruah, A. J. (2025). An Overview of Numerical Simulations in Accretion Physics. *Physics Frontiers*, Vol.-I, pp: 12-17.

[57] Shakura, N. I., & Sunyaev, R. A. (1973). Black holes in binary systems. Observational appearance. *Astronomy and Astrophysics, Vol. 24, p. 337-355*, *24*, 337–355.

[58] Shakura, N. I., & Sunyaev, R. A. (1976). A theory of the instability of disk accretion on to black holes and the variability of binary X-ray sources, galactic nuclei and quasars. *Monthly Notices of the Royal Astronomical Society*, *175*(3), 613–632.

[59] Shapiro, S. L., Lightman, A. P., & Eardley, D. M. (1976). A two-temperature accretion disk model for Cygnus X-1-Structure and spectrum. *Astrophysical Journal, Vol. 204, Feb. 15, 1976, Pt. 1, p. 187-199.*, *204*, 187–199.

[60] Singh, C. B., Okuda, T., & Aktar, R. (2021). Effects of resistivity on standing shocks in low angular momentum flows around black holes. *Research in Astronomy and Astrophysics*, *21*(6), 134.

[61] Singh, M. and Das, S. (2025). Role of thermal conduction in relativistic hot accretion flow around rotating black hole with shock. *Journal of Cosmology and Astroparticle Physics*, vol. 2025, no. 2, Art. no. 068, IOP. doi:10.1088/1475-7516/2025/02/068.

[62] Stoll, M. H., & Kley, W. (2014). Vertical shear instability in accretion disc models with radiation transport. *Astronomy & Astrophysics*, *572*, A77.

[63] Stoll, M. H., Kley, W., & Picogna, G. (2017). Anisotropic hydrodynamic turbulence in accretion disks. *Astronomy & Astrophysics*, *599*, L6.

[64] Svensson, R., & Zdziarski, A. A. (1994). Black hole accretion disks with coronae. *Astrophysical Journal, Part 1 (ISSN 0004-637X), Vol. 436, No. 2, p. 599-606*, *436*, 599–606.

[65] Tanaka, T., & Menou, K. (2006). Hot accretion with conduction: Spontaneous thermal outflows. *The Astrophysical Journal*, *649*(1), 345.

[66] Taverna, R., Marra, L., Bianchi, S., Dovčiak, M., Goosmann, R., Marin, F., Matt, G., & Zhang, W. (2021). Spectral and polarization properties of black hole accretion disc emission: Including absorption effects. *Monthly Notices of the Royal Astronomical Society*, *501*(3), 3393–3405.

[67] Taylor, G. I. (1923), Stability of viscous liquid contained between two rotating cylinders, Phil. *Transi. Roy. Soc. Lond. A*, *223*, 289–343.

[68] Tevzadze, A. G., Chagelishvili, G. D., Zahn, J.-P., Chanishvili, R. G., & Lominadze, J. G. (2003). On hydrodynamic shear turbulence in stratified Keplerian disks: Transient growth of small-scale 3D vortex mode perturbations. *Astronomy & Astrophysics*, *407*(3), 779–786.

[69] Tillmark, N., & Alfredsson, P. H. (1992). Experiments on transition in plane Couette flow. *Journal of Fluid Mechanics*, *235*, 89–102.

[70] Toyouchi, D., Hosokawa, T., Sugimura, K., Nakatani, R., & Kuiper, R. (2019). Super-Eddington accretion of dusty gas on to seed black holes: Metallicity-dependent efficiency of mass growth. *Monthly Notices of the Royal Astronomical Society*, *483*(2), 2031–2043.

[71] Yajima, H., Ricotti, M., Park, K., & Sugimura, K. (2017). Dusty Gas Accretion onto Massive Black Holes and Infrared Diagnosis of the Eddington Ratio. *The Astrophysical Journal*, *846*(1), 3.